\documentclass[11pt,twoside]{article}
\usepackage{FUSE2004}
\usepackage{natbib}

\usepackage{epsf}
\usepackage{psfig}
\usepackage{lscape}
\usepackage{graphicx}
\begin{document}
\title{Properties of Hot, Massive Stars: The Impact of $FUSE$}
\author{Paul A. Crowther}
\affil{Dept of Physics \& Astronomy, University of Sheffield, Hounsfield
Road, Sheffield, S3 7RH, United Kingdom (Paul.Crowther@sheffield.ac.uk)}

\begin{abstract} The impact of $FUSE$ upon the fundamental parameters of
OB stars and Wolf-Rayet stars is reviewed.  The stellar wind signatures
available in the far-UV provide us with important additional diagnostics
of effective temperature. Together with improved non-LTE stellar
atmosphere models allowing for line blanketing and stellar winds, this has
led to a downward revision in the spectral type-temperature calibration
for O stars versus Vacca et al.  (1996) In addition, the Lyman continuum
ionizing fluxes from O dwarfs are compared with previous calibrations of
Panagia (1973) and Vacca et al.  We also discuss mass-loss rates in OB
stars, such that agreement between recent theoretical predictions (Vink et 
al. 2000, 2001) and observations of O supergiants is possible, solely if 
winds are clumped in the far-UV and H$\alpha$ line forming regions, as 
favoured by line profile comparisons for P\,{\sc v} 1118-28 (early to mid 
O) or  S\,{\sc iv} 1062-1073 (late O to early B) in $FUSE$ datasets. In 
contrast,  B supergiant wind strengths are predicted to be much higher 
than observations  indicates, especially if their winds are also clumped. 
Finally,  significant upward revisions in wind velocities of very late WN 
stars are indicated by N\,{\sc ii} 1085 resonance line observations, plus 
elemental abundances in
OB and WR stars are briefly discussed. \end{abstract}

\section{Introduction}

Massive stars ($\geq 10 M_{\odot}$)  play an important role in the ecology
of their host galaxies, which belies their rarity. O stars are generally
the dominant source of Lyman continuum photons in galaxies, whilst their
fast, dense winds are the principal source of kinetic energy in young
bursts of star formation. Wolf-Rayet stars, the chemically evolved
descendants of OB stars, also contribute to chemical and kinetic
enrichment of galaxies via their very dense stellar winds, plus Lyman
continuum photons and ultimately as core-collapse Supernova, which further
dynamically and chemically modify their environment.

Although the majority of our understanding about luminous, hot stars
originates from ground-based optical spectroscopy, their energy
distributions peak in the far- or extreme-UV. The first rocket UV
observations revealed the characteristic P~Cygni signatures of mass-loss
from massive stars (e.g. Morton 1967). Longward of Ly$\alpha$, primarily
$IUE$ has provided observations of a very large database of OB stars,
within the Milky Way, plus Magellanic Cloud supergiants
 (Walborn et al. 1985), supplemented by $HST$ datasets of fainter
Magellanic Cloud stars (Walborn et al. 1995a; Walborn et al. 2000; Evans
et al. 2004b) whilst only a few early-type stars have been observed in the
far-UV prior to $FUSE$, with $Copernicus$ (Snow \& Morton 1976) and more
recently HUT (Walborn et al. 1995b; Schulte-Ladbeck et al. 1995). Far-UV
$FUSE$ atlases have been presented by Walborn et al. (2002) for Magellanic
Cloud OB stars, by Pellerin et al. (2002) for Milky Way OB stars and by
Willis et al. (2004) for Wolf-Rayet stars.

% This present review will summarize our present knowledge of the
% physical properties of OB and Wolf-Rayet stars, focusing especially
% on stellar temperatures and wind properties, with which $FUSE$ has had
% a notable impact. 

\section{Stellar temperatures of early-type stars}

Reliable stellar temperatures ($T_{\ast}$) for early-type stars are of
fundamental importance since bolometric luminosities and ionizing fluxes
are extremely sensitive to temperature. Here we first discuss the
difficulties involved with deriving stellar temperatures for OB stars,
together with the role played by $FUSE$ in identifying the inconsistencies 
from spectroscopic results obtained until recently.

\subsection{Non-LTE models}

Since the UV-optical continuum spectral energy distribution of early-type
stars differ only subtly between O2 and B0 stars, the determination of
$T_{\ast}$ requires the comparison of line profiles of adjacent ionization
stages of the same element (He for O stars, Si for B stars) with model
atmosphere codes. LTE model atmospheres (e.g. Kurucz 1979) are widely used
in stellar astrophysics. However, this treatment assumes that the
ionization state of the gas and the populations of the atomic levels can
be obtained from the local $T_e$ and $n_e$ via the Saha-Boltzmann
distribution, such that collisional processes occur faster than radiative
processes.  Unfortunately, the opposite is true in hot star winds, so it
is necessary to solve the equation of statistical equilibrium everywhere,
i.e. non-LTE.

The major complication of non-LTE is that a determination of populations
uses rates which are functions of the radiation field, itself is a
function of the populations. Consequently, it is necessary to solve for
the radiation field and populations simultaneously, which is
computationally demanding, and requires an numerical iterative scheme to
obtain consistency. Unfortunately, the problem is too complex for
analytical solutions.  Considerable effort has gone into developing
realistic non-LTE model atmospheres for early-type stars in recent years
by a number of independent groups.  Up until very recently, the non-LTE
model atmosphere code TLUSTY (Hubeny 1988) or SURFACE/DETAIL (Butler \&
Giddings 1985) represented the standard approach for O stars, based upon
H-He plane-parallel geometry. A compilation of properties of O stars based
largely upon such an approach was presented by Vacca et al. (1996).

However, two potentially significant effects were lacking in this
approach; (i) the atmospheres of early-type stars are not composed of
solely H and He, such that the collective effect of metals within the
atmosphere may significantly affect the ionization structure via so-called
line blanketing. Unfortunately, consistently treating metal `line
blanketing' in non-LTE atmospheres is computationally demanding, although
great progress has now been achieved by various groups including TLUSTY
(Hubeny \& Lanz 1995; Lanz \& Hubeny 2003);  (ii) the presence of stellar
winds complicates the geometry, such that a proper treatment requires
spherical geometry. Winds may contaminate (`fill-in') photospheric
absorption lines in OB stars, causing temperatures to be overestimated
still further in the standard plane-parallel assumption (Schaerer \&
Schmutz 1994; Bohannan et al. 1990).

At present, there are two line blanketed non-LTE spherical model
atmospheres widely in use that permit (consistent) detailed studies of
far-UV, UV and optical spectroscopy of O stars, namely CMFGEN (Hillier \&
Miller 1998)  and Fastwind (Santaloya-Rey et al. 1997; Herrero et al.
2002).  In addition, WM-basic (Pauldrach et al. 2001) permits UV spectral
synthesis, although presently this lacks an appropriate treatment of line
broadening for optical spectral features.

\subsection{Temperature scale of early-type stars}

A number of quantitative studies of OB stars were carried out in the last
decade in which non-LTE plane-parallel model atmospheres for photospheric
optical lines of H and He (to derive $T_{\ast}$) were successfully
combined with spectral synthesis codes for UV ($IUE/HST$) metal lines (to
derive wind properties), such as Pauldrach et al. (2001) and Haser et al.
(1998). However, Early Release Observations from $FUSE$ included two mid-O
supergiants for which previously optically derived temperatures proved
very poor matches to the UV spectral region (Fullerton et al. 2000).  In
contrast with the familiar N\,{\sc v}, Si\,{\sc iv} and C\,{\sc iv}
saturated P~Cygni line profiles from $IUE/HST$, unsaturated lines from
trace ions were covered by $FUSE$, e.g. C\,{\sc iii}, N\,{\sc iii},
S\,{\sc iv}, P\,{\sc iv} permitting a greater diagnostic role with regard
to the ionization balance of metal species. For the case of Sk~80 (AzV
232, O7Iaf$^{+}$) in the SMC, the metal lines in the $FUSE$ spectral
window indicated a substantially lower $T_{\ast}$ (by 15\%) than previous
helium lines from optical studies (Puls et al. 1996).

\begin{figure}[htbp!]
\plotfiddle{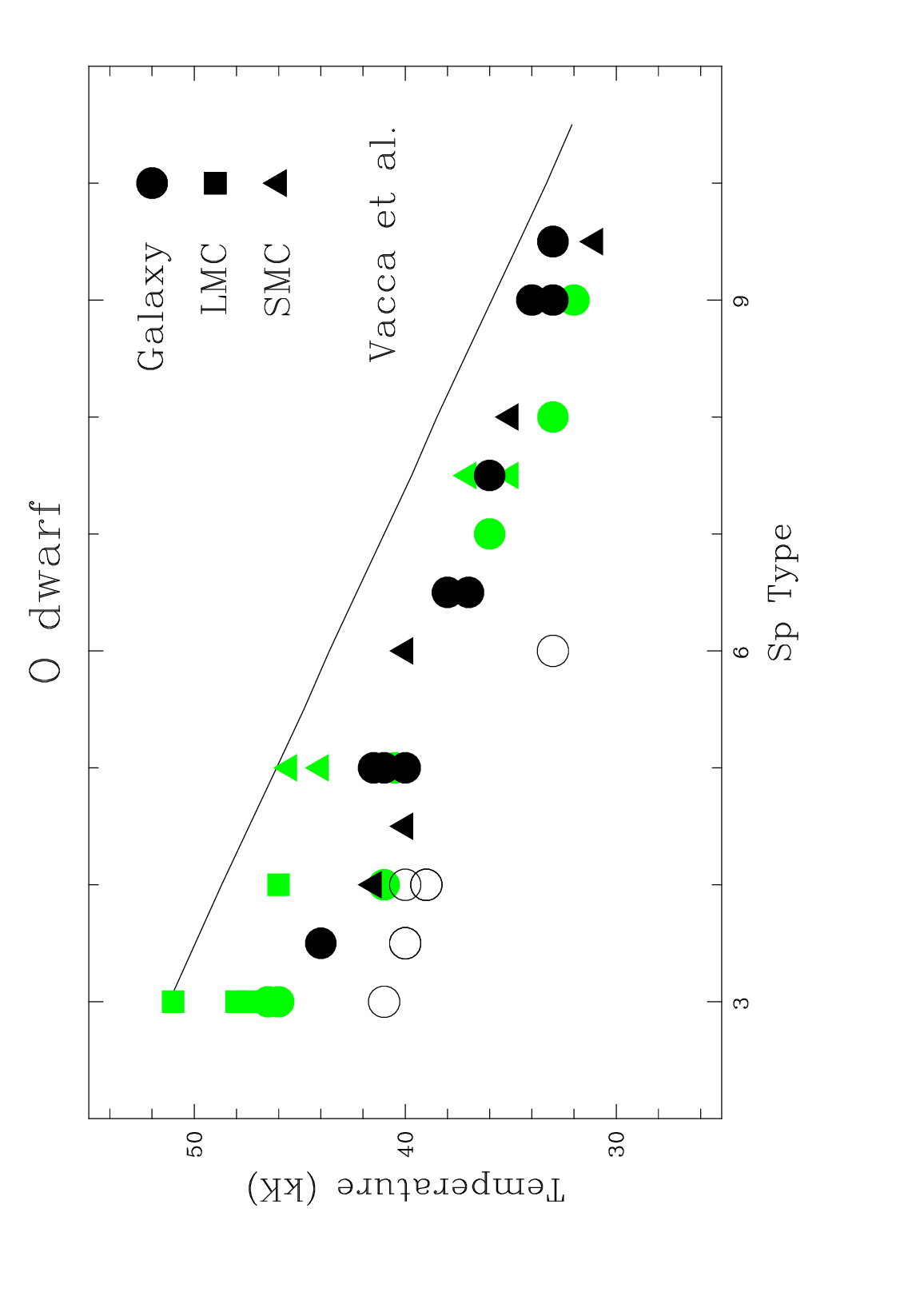}{7.7cm}{-90}{50}{50}{-215}{+235}
\caption{Effective temperatures of O dwarfs versus the Vacca
et al. (1996) calibration, based on Galactic results from
Martins et al. (2004), Bianchi \& Garcia (2002), Garcia \& Bianchi 
(2004) and Repolust et al. (2004), 
LMC/SMC results from Massey et al. (2004) and Bouret et al. (2003).
For this and subsequent figures, CMFGEN (black filled), FASTWIND 
(grey filled) and WM-basic (open).}\label{fig_odwarf}
\plotfiddle{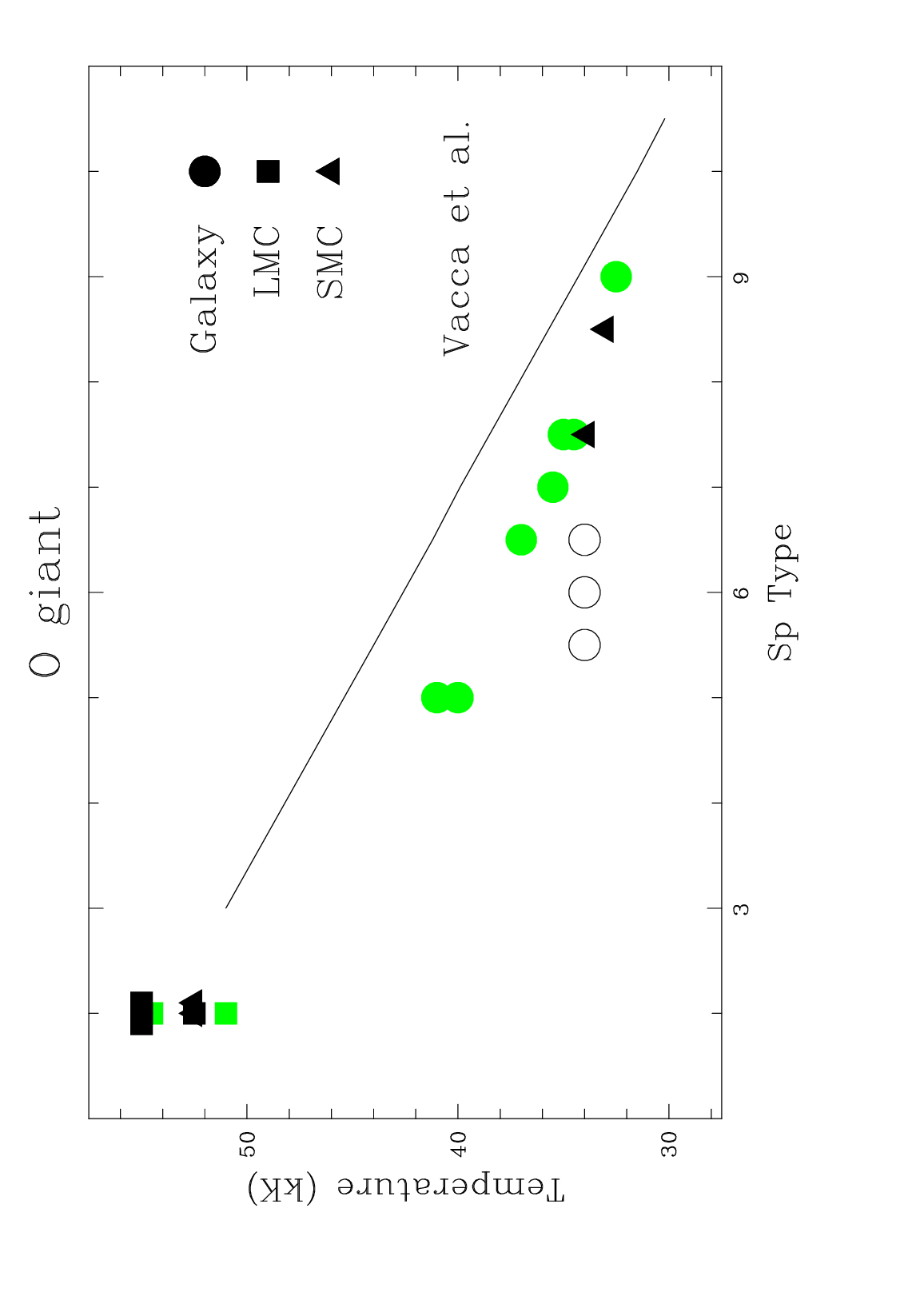}{7.7cm}{-90}{50}{50}{-215}{+235}
\caption{Effective temperatures of O giants and bright giants
versus the Vacca et al. (1996) calibration, based on Galactic results from 
Herrero et al. (2002), Bianchi \& Garcia
(2002) and Repolust et al. (2004),  LMC/SMC results 
from Bouret et al. (2003), Walborn et al. (2004),
Hillier et al. (2003), Massey et al. (2004) and Evans et al. (2004)}
\label{fig_ogiant}
\end{figure}

\begin{figure}[t]
\plotfiddle{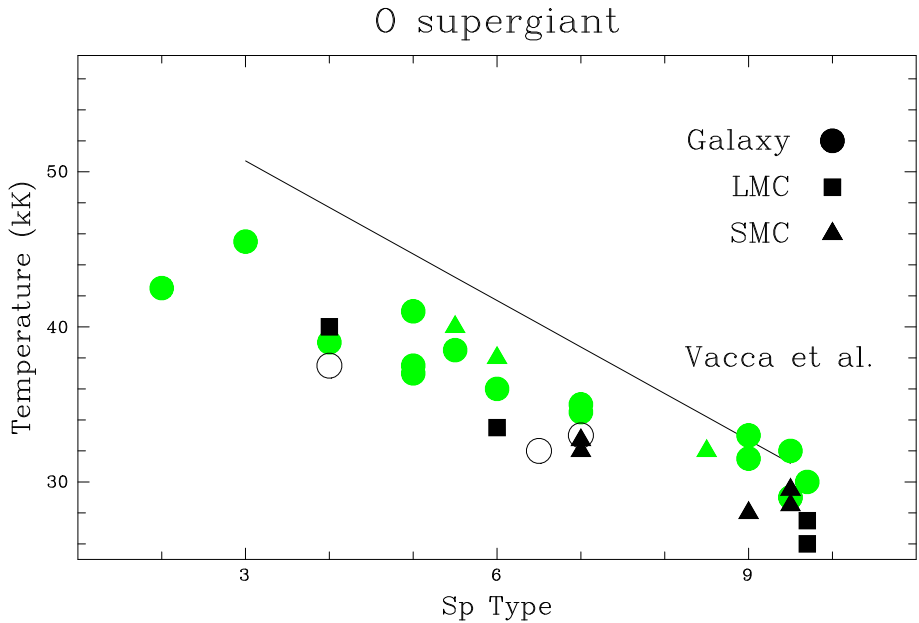}{7.7cm}{-90}{50}{50}{-215}{+235}
\caption{Effective temperatures of O supergiants versus the Vacca et al.
(1996) calibration, based on Galactic results from Herrero et al.
(2002), Bianchi \& Garcia (2002), Repolust et al. (2004) and Garcia \& Bianchi (2004), 
LMC/SMC results from Crowther et al. (2002), Hillier et al. (2003), 
Evans et al. (2004), Massey  et al. (2004)}
\label{fig_osuper}
\end{figure}

This major inconsistency was ultimately resolved via the use of line 
blanketed, 
spherical, non-LTE models, introduced above, together with
$FUSE$, $IUE/HST$ and optical datasets
in which the wind and line blanketing had a substantial
effect on the helium ionization balance (see e.g. Repolust et al.
2004), and 
confirmed the earlier $FUSE$ far-UV result for Sk~80 (Crowther et al. 
2002a).
Subsequently there have been a number of studies of Galactic and
Magellanic Cloud O stars based on Fastwind and CMFGEN analyses of optical
datasets, plus UV/far-UV spectroscopy in some instances. Overall,
recent results indicate a lower $T_{\ast}$ scale for
dwarfs, giants and supergiants, as indicated in Figs.~\ref{fig_odwarf},
\ref{fig_ogiant} and \ref{fig_osuper}, respectively. 
The trend towards 2--3kK lower $T_{\ast}$ is also true for
B supergiants (Crowther et al. 2005).

Agreement between results from the different model atmosphere codes is
good in all cases for which optical diagnostics are included, across
each host galaxy.  The principal outliers are results from Bianchi 
\& Garcia (2002) and Garcia \& Bianchi (2004)  in which solely far-UV 
$FUSE$ and UV $IUE$ spectroscopy of Galactic O stars were analysed using 
WM-basic. Since it is not possible to compare their predicted optical 
photospheric lines with observations, caution is presently advised 
regarding their 
validity.

In contrast with O and early B stars, recent spectroscopic results for
Wolf-Rayet stars have, in general, led to an increase in stellar
temperatures\footnote{Since WR winds are optically thick, stellar
temperatures generally refer to deep layers (Rosseland optical depth
$\tau_{\rm Ross}\sim$10 or 20) rather than the conventional $\tau_{\rm
Ross}$=2/3} relative to earlier calculations. Current line blanketed
models for WR stars (e.g. CMFGEN)  are effectively identical to those used
for O stars with winds, whilst previous studies employed non-LTE H-He or
H-He-CNO spherical non-LTE models. Why has the temperature scale for WR
stars moved in the opposite sense? Blanketing has the effect of
re-distributing extreme-UV flux to longer wavelengths such that higher
temperatures are required to maintain a specific {\it wind} ionization
balance of, say, helium for WN stars or carbon for WC stars.

\subsection{Ionizing fluxes}

\begin{figure}[tb!]
\plotfiddle{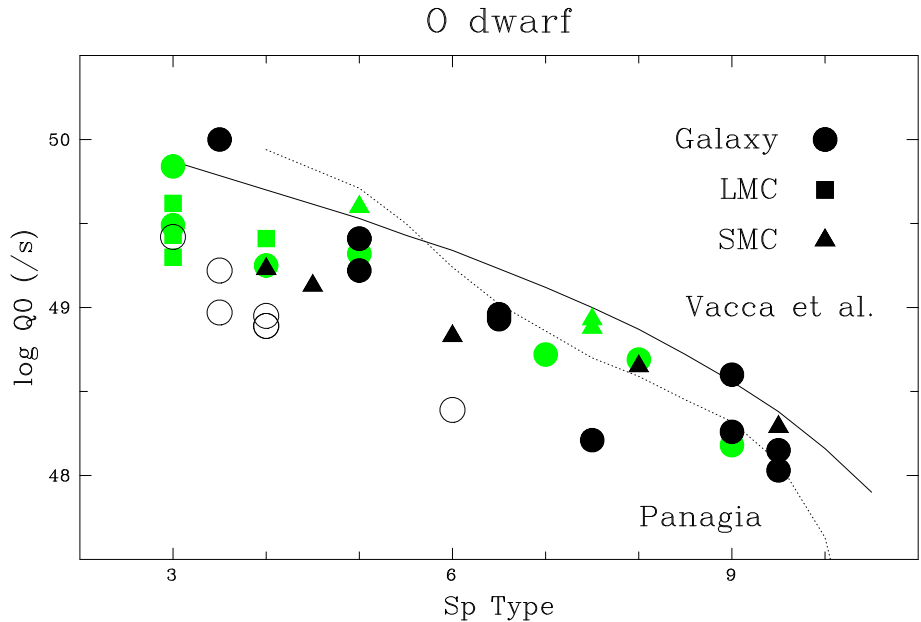}{7.7cm}{-90}{50}{50}{-215}{+235}
\caption{Lyman continuum fluxes for O dwarfs versus calibrations of 
Panagia (1973, dotted),
Vacca et al. (1996, solid) applying
Table~4 from Lanz \&  Hubeny (2003)  to results indicated in Fig.1.}
\label{fig_lyman}
\end{figure}

Naturally, lower $T_{\ast}$ for OB stars impacts upon bolometric
luminosities (since bolometric corrections are very sensitive to
temperature for O stars) and Lyman continuum fluxes, due to lower, softer
extreme-UV fluxes.  By way of example, in Fig.~\ref{fig_lyman} we compare
recent results for O dwarfs discussed above with calibrations from Panagia
(1973) and Vacca et al. (1996). We find that the Vacca et al. calibration
overestimates contemporary determinations of ionizing fluxes from O stars 
at all spectral types, whilst the Panagia calibration is more reliable at 
later subtypes, but also overestimates fluxes amongst early O stars. Such 
revisions naturally have great influence on the stellar content of H\,{\sc 
ii} regions, although the widely adopted N(LyC)=10$^{49}$ ph s$^{-1}$ for 
O7V stars remains a reasonable assumption (Vacca 1994). Conversely, higher 
stellar temperatures of WR stars leads to a greater contribution from such 
stars in the Lyman continuum ionization budget of H\,{\sc ii} regions.

\section{Wind properties of early-type stars}

Global wind properties of hot stars may be characterized by mass-loss and
wind velocity. The former depends on application of varying complexity of
theoretical interpretation, whilst the latter can be directly measured
with minimal interpretation.

\subsection{Wind velocities}

UV and far-UV P~Cygni profiles from metal resonance transitions provide a
direct indication of stellar wind velocities of early-type stars. Two
approaches are possible -- either the maximum blueward extent of saturated
`black' absorption troughs can be directly measured from observations
(e.g. Prinja \& Crowther 1998), or fit using techniques such as the
Sobolev with exact integration (SEI) method (e.g. Haser et al. 1998).
Since many far-UV P~Cygni lines are unsaturated, 'black' wind velocity
measurements from $FUSE$ spectroscopy are generally restricted to a few
strong wind lines, such as N\,{\sc iii} $\lambda$989--991. Willis et al.
(2004) compare wind velocities of Wolf-Rayet stars from $FUSE$ datasets
with literature UV/optical measurements, and find overall good agreement,
with the exception of very late subtype WN stars, for which N\,{\sc ii}
$\lambda$1085 P~Cygni profiles indicated wind velocities up to a factor of
two times higher than previously estimated.

\subsection{Mass-loss rates -- evidence for clumping?}

Several techniques are available for the determination of empirical
mass-loss rates in early-type stars, involving radio, optical and
UV/far-UV observations. Theoretical mass-loss rates have been published by
Vink et al. (2000, 2001).

\subsubsection{Radio}

For nearby Galactic OB stars with strong winds, probably the most robust
method of determining mass-loss rates is via the thermal free-free excess
at IR/mm/radio wavelengths following e.g. Wright \& Barlow (1975).
Unfortunately, stars with relatively weak winds possess very modest radio
excesses, multiple frequency observations are necessary to ensure against
non-thermal radio emission from colliding winds, plus no progress with
extragalactic early-type stars is presently possible due to sensitivity
limits with current facilities. The radio photosphere of O stars (hundreds
of $R_{\ast}$) greatly exceeds that from UV/optical emission lines
(typically 1--2 $R_{\ast}$).

\subsubsection{Optical}

Optical (e.g. H$\alpha$) or near-IR (e.g. Br$\alpha$) spectroscopy offers
a readily available indicator of mass-loss in early-type stars within the
Local Group, subject to difficulties with nebular contamination from
H\,{\sc ii} regions (Massey et al. 2004).  The main limitation with such
diagnostics are that complex non-LTE models introduced above need to be
employed for reliable mass-loss rates (e.g. Repolust et al. 2004).
Alternatively, analytical techniques can be used (e.g, Puls et al. 1996),
provided they are suitably calibrated against non-LTE model results
(Markova et al. 2004). 

To date, the most extensive multi-wavelength studies of mass-loss,
specifically, Crowther et al. (2002), Hillier et al.  (2003) and Evans et
al. (2004) have derived mass-loss rates of Magellanic Cloud OB stars from
H$\alpha$ observations, and compared predicted UV and far-UV line profiles
with observations. With one exception (AzV~235), agreement was very good
except that the predicted P~Cygni absorption components of P\,{\sc v}
$\lambda$1118-28 (in mid O stars) and S\,{\sc iv} $\lambda$1062-1068 (in
late O/early B stars) were too strong. One was able to resolve the
discrepancy either by (a) reducing the Phosphorus abundance below that
expected (although the Sulphur abundance is well known from nebular
studies) or, (b) introducing clumped winds in which lower mass-loss rates
were able to reproduce the H$\alpha$ profile.  In general, clumped versus
homogeneous models showed otherwise very subtle differences, such that the
availability of unsaturated P~Cygni profiles in $FUSE$ datasets offers the
best opportunity to investigate clumping in OB stars.

\subsubsection{UV and far-UV}

Finally, the SEI method can be used to derive optical depths (or
alternatively $\dot{M} q$, where $q$ represents the ionization fraction of
the specific ion relative to the total for that element) versus velocity
for unsaturated UV and far-UV P~Cygni line profiles. Massa et al. (2003)
have analysed a sample of LMC O stars using the SEI method. Rather than
derive mass-loss rates, they instead adopt theoretical mass-loss rates
from Vink et al. (2000) to investigate the ionization balance in O stars.
From P\,{\sc v} $\lambda$1118-28 they find $q$(P$^{5+}$) never exceeds
$\sim$0.2, such that either (i) the calculated mass-loss rates are too
high; (ii) the adopted P abundance is too large, or (iii) the winds are
strongly clumped. The latter possibility obviously ties in naturally with
the above arguments from more complex techniques. Massa (these
proceedings) discusses recent extensions to this study involving Galactic
OB stars.

\begin{figure}[ht!]
\plotfiddle{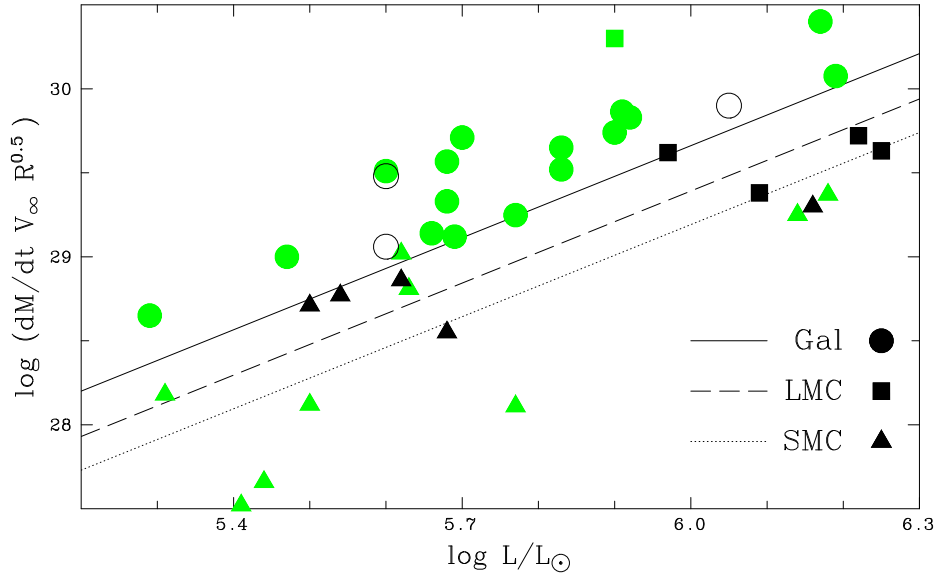}{7.5cm}{-90}{50}{50}{-210}{+230}
\caption{Comparison between wind momenta of O2--O9.7 
supergiants in the Milky Way (Repolust et al. 2004; Herrero et al. 2002), 
with LMC and SMC  counterparts (Crowther et al. 2002; Evans et al. 2004; 
Massey et al. 2004),  based on homogeneous mass-loss rates. 
For comparison we show the theoretical Solar metallicity calibration by 
Vink et al. (2000) for $T_{\rm eff} \geq 27.5$kK (solid line),
plus corresponding predictions for 0.4$Z_{\odot}$ (LMC, dashed) and
0.2$Z_{\odot}$ (SMC, dotted) following Vink et al (2001).
}\label{Owind}
\end{figure}

\subsection{An empirical metallicity dependence of wind strength?}

Current radiatively driven wind theory predicts a modest dependence of
mass-loss rate versus metallicity (Z, Vink et al. 2001). Consequently,
considerable effort is presently underway to establishing an empirical
dependence by comparing Solar neighbourhood OB stars to counterparts
in the LMC and SMC, where the metallicity is a factor of $\sim$2 to
$\sim$4--5 times lower, respectively.  

\begin{figure}[ht!]
\plotfiddle{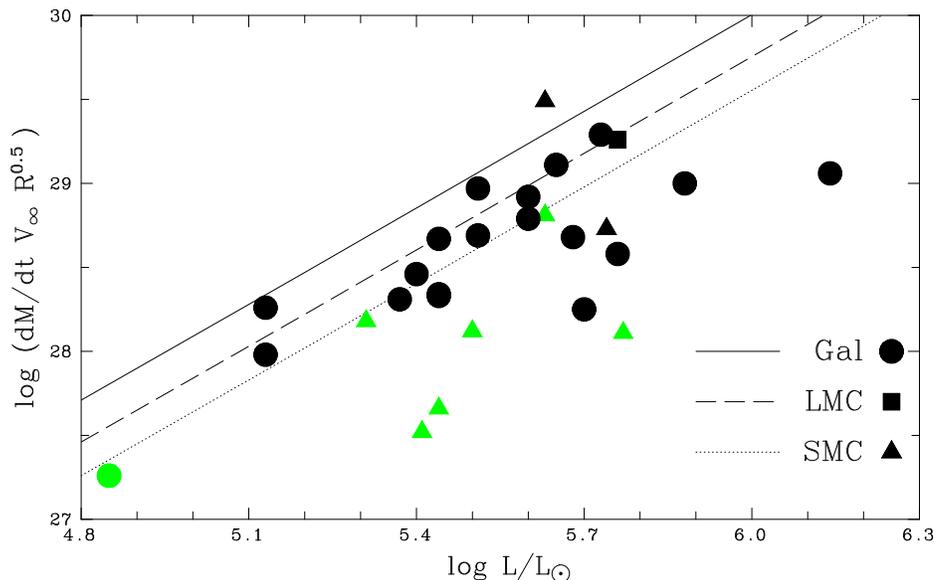}{7.5cm}{-90}{50}{50}{-210}{+230}
\caption{Comparison between wind momenta of B0--B3 
supergiants in the Milky Way (Crowther et al. 2005; Herrero et al. 2002), 
with  LMC and SMC  counterparts (Evans et al. 2004; Trundle et al. 2004), 
again based on homogeneous mass-loss  rates. For comparison we show the 
theoretical Solar metallicity  predictions by 
Vink et al. (2000) for $T_{\rm eff} \leq 22.5$kK (solid line).
plus corresponding predictions for 0.4$Z_{\odot}$ (LMC, dashed) and
0.2$Z_{\odot}$ (SMC, dotted) following Vink et al (2001).
}\label{Bwind}
\end{figure}

Wind velocities of LMC  O stars differ little from Galactic counterparts, 
although a  more prominent effect is observed in the SMC amongst early O 
stars (Walborn et al. 1995a; Prinja \ Crowther 1998). 
More critically, in Fig.~\ref{Owind} we present derived wind momenta of O 
supergiants in the  Milky Way/LMC/SMC, {\it assuming} homogeneous winds, 
 with  the Vink et al. (2000) Galactic 
calibration (for 
$T_{\ast}  \geq$27.5kK), together with predictions for LMC and SMC 
metallicities ($\dot{M} \propto  Z^{0.69}$: Vink et al. 2001).
The comparison appears poor, although modest clumping would resolve the 
Galactic O  supergiant  mismatch, via a decrease in empirical mass-loss 
rates by $\sim$0.3--0.5 dex. For the SMC supergiants there is a greater
observational scatter, such that some supergiants would agree with the
prediction for clumped winds, whilst others would fall far below.
A similar comparison for B supergiants is presented in 
Fig.\ref{Bwind}, now relative to the Vink et al. calibtrations for stars 
with $T_{\rm eff}\leq$22.5kK ( $\dot{M} \propto  Z^{0.64}$: Vink et al. 
2001). In contrast, the measured wind densities of Galactic and SMC 
B  supergiants  fall below the predictions, such that clumping  would only 
worsen the comparison. Consequently, significant problems clearly remain, 
although observationally SMC supergiants do show weaker wind densities 
than LMC  and Galactic counterparts, although a larger  sample size for 
all galaxies  is urgently required for firmer conclusions. Such a study is 
ongoing via  several groups, including our own. Our approach is to  
combine high  quality  optical (VLT/UVES) spectroscopy with far-UV 
($FUSE$)  spectroscopy of  Magellanic Cloud OB stars drawn from the 
Guaranteed Time program P117 (P.I.: J. Hutchings) plus the ongoing Legacy 
program P511 (P.I.: W. Blair).

\subsection{Wolf-Rayet stars}

Finally, let us briefly discuss WR stars. Since their winds 
are denser than OB stars, they offer  a great
many more wind diagnostics, although as their photospheres are not 
directly  observable, it is necessary to derive their wind and physical
properties simultaneously. Crowther et al. (2000) illustrates the quality
of synthetic fits to far-UV/UV/optical/near-IR spectroscopy that is possible,
for the case of a WO star in the LMC. WR winds have long been established
to be clumped, specifically via the strength of electron scattering wings
on line profiles (Hillier 1991), rather than far-UV P~Cygni profiles.

In contrast with OB stars, historically no (heavy)
metallicity dependence of WR winds has been adopted, although
Crowther et al. (2002b) argue in favour of a metallicity dependence,
on the basis of iron-peak lines providing the principal line driving 
in WR stars  (even within the carbon-rich atmosphere of a WC star).
C\,{\sc iii} $\lambda$5696 was identified as particularly 
sensitive to wind strength, such that early WC subtypes (with weak/absent 
$\lambda$5696) are anticipated
in metal poor environments such as the LMC, as observed (Breysacher et al. 
1999), and late WC subtypes (with strong $\lambda$5696) are anticipated in 
metal rich environments, such as M83, also as observed (Crowther et al. 
2004).

\section{Elemental Abundances}

To date, the bulk of stellar abundance studies of OB-type stars
have focussed upon H/He contents (e.g. Herrero et al. 1992) rather
than light (CNO) or heavy (iron peak) elements. Despite numerous metal
lines in the $FUSE$ spectral window, these are primarily wind influenced,
such that the prime abundance lines are observed in the blue and yellow 
visible. Recent studies of OB supergiants by Crowther et al. (2002a),
Hillier et al. (2003) and
Evans et al. (2004a) reveal partially CNO-cycle processed material
at their surfaces. Indeed, Walborn et al. (2004) find similar levels
of nitrogen enhancement and oxygen depletion in a subset of early O giants.
Iron abundances of Magellanic Cloud O stars
have been estimated by Haser et al. (1998) from UV $HST$/FOS spectroscopy,
and by Hillier et al. (2003) from $HST$/STIS spectroscopy. 
Since the bulk of the iron lines (Fe\,{\sc iv--vi} observed in O stars 
fall in the  $HST/IUE$ domain, no further progress has been possible with 
$FUSE$.

Metal abundance studies of Wolf-Rayet stars are also optically derived,
in general, although iron lines are again exclusive to the $HST/IUE$
region, and suitable oxygen lines in WC stars are located around 
$\lambda$3000\AA\ (Hillier \& Miller 1999). Overall, WC and especially
WO stars are very rich in carbon and oxygen (Crowther et al. 2000, 2002b)
with hydrogen and nitrogen absent, whilst fully processed CNO cycle
products are observed in WN stars, with hydrogen present in most late
WN subtypes (Crowther et al. 1995). 
The most comprehensive abundance studies of WR stars to date 
have been carried out for two late WN stars by Herald et al. (2001), 
including  $HUT$ spectroscopy. Detailed  studies using 
$FUSE$ datasets are presently ongoing.

%\section{Summary}

%We review our present knowledge of the temperatures, wind properties
%and abundances of OB and Wolf-Rayet stars, with particular attention
%paid to the contribution of $FUSE$, namely the presence of metal P~Cygni
%resonance lines that provide (i) constraints on the optically
%derived temperatures of O stars, (ii) indicators of wind clumping via
%from both non-LTE synthetic models and SEI line profile fits, (iii) 
%additional
%checks on wind velocities from early-type stars, such as 
%N\,{\sc ii} $\lambda$1085 in late WN stars. 

\acknowledgements Thanks to Alex Fullerton for maintaining an excellent
database of early-type $FUSE$ observations, and to 
Fabrice Martins for providing recent results
on Galactic O dwarfs prior to publication. Financial support is provided
by the Royal Society.

% If you wish to use BiBTeX uncomment and fill in the .bib file name.  Note that 
% we are still (July 23) waiting for input from the ASP as to which 
% "bibliographystyle" to use.  "natbib" is unlikely to be the right one, but is 
% left here as a place holder.

%\bibliography{bib-file}
%\bibliographystyle{natbib}

% For using the "thebibliography" environment use these.
% See the "Authors Instructions" for details.

\end{document}